\documentclass[apex]{jjap3}
\usepackage{txfonts}
\usepackage{color}   
\usepackage{graphicx} 
\usepackage{subfigure}
\usepackage{amsmath}

\title{Entanglement generation using a controlled-phase gate for time-bin qubits}
\author{Hsin-Pin Lo\thanks{E-mail: hsinpin.lo@lab.ntt.co.jp} , Takuya Ikuta, Nobuyuki Matsuda, Toshimori Honjo, and Hiroki Takesue}
\inst{NTT Basic Research Laboratories, NTT Corporation, 3-1 Morinosato Wakamiya, Atsugi, Kanagawa, 243-0198, Japan}
\abst{Quantum logic gates are important for quantum computations and quantum information processing in numerous physical systems. While time-bin qubits  are suited for quantum communications over optical fiber, many essential quantum logic gates for them have not yet been realized. Here, we demonstrated a controlled-phase (C-Phase) gate for time-bin qubits that uses a 2x2 optical switch based on an electro-optic modulator. A Hong-Ou-Mandel interference measurement showed that the switch could work as a time-dependent beam splitter with a variable spitting ratio. We confirmed that two independent time-bin qubits were entangled as a result of the C-Phase gate operation with the switch.} 

\begin{document}
\maketitle

As the progress of digital computers has begun to reach its limits, expectations regarding quantum computation are rapidly growing. Many physical implementations of quantum computers have been demonstrated using various quantum bit (qubit) systems such as superconducting devices \cite{QST_superconducting_IBM, cnot_gate_superconducting, gate_superconducting}, ion traps \cite{steane_iontrap, gate_ion_trap}, and quantum dots \cite{Li_Qdot_gate, Hanson_dot_RMP_2007, Zwanenburg_dot_RMP_2007}. An essential research topic here is building quantum logic gates with which are implemented interactions among qubits to form multi-partite entangled states \cite{Nielsen_Chuang}. It is difficult to realize such photon interactions with linear optics. Despite this, there are proposed protocols for realizing universal quantum computers only with linear optical tools \cite{KLM, cnot_theory, cphase_theory, GHZ_LOQC_2015}. In fact, quantum logic gates for photonic qubits have been demonstrated using path \cite{gate_photon_OBrien} and polarization qubits \cite{cnot_Pittman_2003, cphase_pdbs, cnot, gate_Bell, kok_RMP}. Such quantum logic gates are also useful quantum communications, in which quantum error correction using simple quantum logic gates for photonic qubits enables a quantum repeater system without using quantum memories based on matter qubits \cite{Bill_repeater}. 

Most of the quantum communication systems experimented with to date over optical fiber have difficulty using these types of photonic qubits, because it is difficult to preserve the relative phase between the modes that compose the qubits as a result of fluctuation in the refractive indices and birefringence of the optical fibers. Instead, time-bin qubits have been used in most of the fiber-based quantum communication experiments because of their robustness against these fluctuations \cite{time-energy_Brendel_1999, time-energy_Gisin_2007, Honjo_timebin_QKD}. However, a problem remains in that many of the essential quantum logic gates, such as controlled-not (CNOT) or controlled-phase (C-Phase) gates, have not yet been realized for time-bin qubits.

In this study, we demonstrate an implementation of the C-Phase gate for time-bin qubits. Our scheme is based on a two-input, two-output (2x2) optical switch used as a time-dependent beam splitter \cite{takesue_switch}. We show that two time-bin qubits can be entangled as a result of C-Phase gate operation with the 2x2 switch.

\begin{figure*}[t]
\centering
\includegraphics[width=17.5cm]{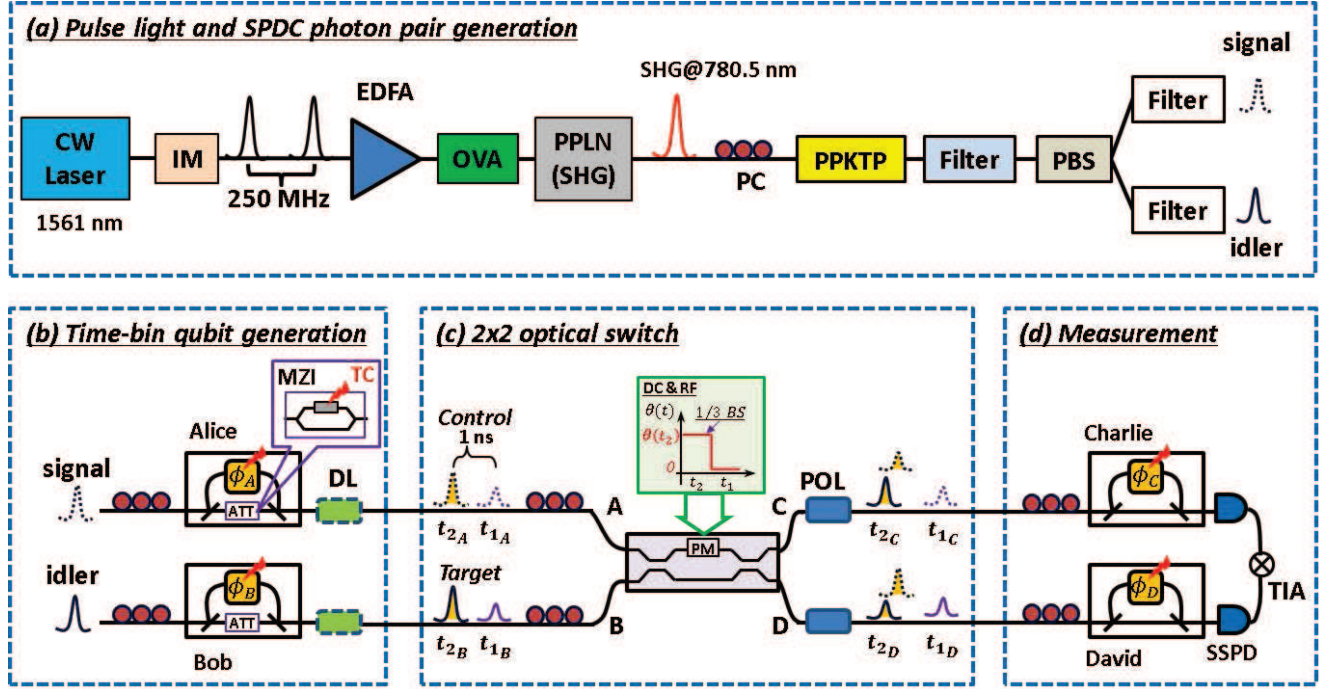}
\caption{Experimental setup. CW Laser: continuous wave laser. IM: intensity modulator. EDFA: erbium-doped fiber amplifier. OVA: optical variable attenuator. PPLN: periodically poled lithium niobate waveguide. SHG: second harmonic generation. PC: polarization controller. PPKTP: periodically poled potassium titanyl phosphate waveguide. PBS: polarization beam splitter. Filter: bandpass filter. ATT: amplitude attenuator to the first temporal mode. MZI: Mach-Zehnder interferometer. TC: temperature controller. PM: electro-optic phase modulator. DL: optical delay line. POL: polarizer. SSPD: superconductor single-photon detector. TIA: time interval analyzer.}
\label{expsetup}
\end{figure*}

Here, we describe the C-Phase gate operation for time-bin qubits by using a high-speed 2x2 optical switch, which is a Mach-Zehnder (MZ) interferometer that includes an electro-optic phase modulator (PM) \cite{takesue_switch} in one of the optical paths as shown in Fig.~\ref{expsetup}(c). We launch two time-bin qubits as a control and a target state into ports A and B of the switch, whose states are given by $\left|\psi \right\rangle_A = c_{1A} \left|t_1 \right\rangle_A + c_{2A} e^{i \phi_A} \left|t_2 \right\rangle_A$ and $\left|\psi \right\rangle_B = c_{1B} \left|t_1 \right\rangle_B +c_{2B} e^{i \phi_B} \left|t_2 \right\rangle_B$. The index A and B are the input ports to Alice and Bob. Here, $\left|t_{x}\right\rangle_y$ represents the photon in the time position $t_x \in \{t_1, t_2\}$ of the input port, $y \in \{A, B\}$, and $c_{xy}$ is the amplitude of $\left|t_{x}\right\rangle_y$ which is a nonnegative real number that satisfies $c_{1y}^2+c_{2y}^2=1$, and $\phi_y$ is the phase difference between temporal states $t_1$ and $t_2$ which can be set by adjusting the temperature controller (TC). The ideal C-Phase gate operates on two input time-bin states, as follows:
\begin{eqnarray}
\begin{split}
\left|\psi_{in} \right\rangle =& \left|\psi \right\rangle_A \otimes \left|\psi \right\rangle_B \\
=& c_{1A} c_{1B} \left|t_1 \right\rangle_A \left|t_1 \right\rangle_B + c_{1A} c_{2B} e^{i \phi_B} \left|t_1 \right\rangle_A \left|t_2 \right\rangle_B \\
+& c_{2A} c_{1B} e^{i \phi_A} \left|t_2 \right\rangle_A \left|t_1 \right\rangle_B +  c_{2A} c_{2B} e^{i(\phi_A+\phi_B)} \left|t_2 \right\rangle_A \left|t_2 \right\rangle_B.
\end{split}
\end{eqnarray}
By applying a time-varying signal to the PM, the 2x2 switch can work as a time-dependent beam splitter whose splitting ratio changes in time. The evolution of a time-bin state with the 2x2 switch is described as
\begin{eqnarray}
\begin{split}
\left|t_k \right\rangle_A&=\cos\left(\frac{\theta(t_k)}{2}\right)\left|t_k \right\rangle_C-\sin\left(\frac{\theta(t_k)}{2}\right)\left|t_k \right\rangle_D, \\
\left|t_k \right\rangle_B&=\sin\left(\frac{\theta(t_k)}{2}\right)\left|t_k \right\rangle_C+\cos\left(\frac{\theta(t_k)}{2}\right)\left|t_k \right\rangle_D, 
\label{sw1}
\end{split}
\end{eqnarray}
where $\theta(t_k)$ represents the phase difference between the two arms of the MZ interferometer at time $t_k$ and the index C and D are the output ports to Charlie and David. For the C-Phase gate operation, we set $\theta(t_1)=0$ and $\theta(t_2)=2\cos^{-1}(\frac{1}{\sqrt{3}})$, which means that the 2x2 switch passes the first temporal mode and works as a one-third beam splitter for the second mode.

By performing a coincidence measurement between Charlie and David, we obtain a state given by
\begin{eqnarray}
\begin{split}
&c_{1A}c_{1B} \left|t_1 \right\rangle_C \left|t_1 \right\rangle_D + c_{1A}c_{2B}e^{i\phi_B}\sqrt{\frac{1}{3}}\left|t_1 \right\rangle_C \left|t_2 \right\rangle_D  \\
+ & c_{2A}c_{1B}e^{i\phi_A} \sqrt{\frac{1}{3}}\left|t_2 \right\rangle_C  \left|t_1 \right\rangle_D - c_{2A}c_{2B}e^{i(\phi_A+\phi_B)} \frac{1}{3}\left|t_2 \right\rangle_C \left|t_2 \right\rangle_D. 
\end{split}
\end{eqnarray}
Similarly to the case of previous C-Phase gates realized for path \cite{gate_photon_OBrien} and polarization \cite{cphase_theory, cphase_pdbs} qubits, the amplitude unbalance can be eliminated by applying one-third attenuation only to the $t_1$ mode. Thus, in the coincidence basis between Charlie and David, we obtain an output state for the C-Phase gate operation, given by
\begin{eqnarray}
\begin{split}
\left|\psi_{out} \right\rangle =& c_{1A} c_{1B} \left|t_1 \right\rangle_C \left|t_1 \right\rangle_D + c_{1A} c_{2B} e^{i \phi_B} \left|t_1 \right\rangle_C \left|t_2 \right\rangle_D \\
+& c_{2A} c_{1B} e^{i \phi_A} \left|t_2 \right\rangle_C \left|t_1 \right\rangle_D -  c_{2A} c_{2B} e^{i(\phi_A+\phi_B)} \left|t_2 \right\rangle_C \left|t_2 \right\rangle_D.
\end{split}
\end{eqnarray}

The experimental setup is shown in Fig.~\ref{expsetup}. We generate a 1561-nm pulse train with a 250-MHz repetition rate by modulating continuous-wave laser light from an external-cavity diode laser using a lithium-niobate intensity modulator (IM). The pulsed light is amplified by an erbium-doped fiber amplifier (EDFA). The amplified pulse train is launched into a periodically poled lithium niobate (PPLN) waveguide to generate a 780.5-nm pulse train via second harmonic generation (SHG). Then, the SHG light pumps the type-II periodically poled potassium titanyl phosphate (PPKTP) waveguide to generate quantum-correlated photon pairs through the process of spontaneous parametric down-conversion (SPDC). The band-pass filter placed after the PPKTP waveguide is used to block the SHG pump light. By changing the optical variable attenuator (OVA), we adjust the average power of pump light to set the average number of correlated photon pairs per pulse at 0.028. The photon pairs are input into a polarization beam splitter (PBS) to separate the signal and idler photons for Alice and Bob. Then the photons are passed through another band-pass filter with a central wavelength of 1561 nm and a bandwidth of 1.4 nm to reduce the noise, as shown in Fig.~\ref{expsetup}(a).  

Alice and Bob prepare their time-bin states by launching the signal and idler photons into 1-bit delay interferometers fabricated using planar light-wave circuit (PLC) \cite{takesue_plc} technologies, as shown in Fig.~\ref{expsetup}(b). As discussed in the previous section, one-third amplitude attenuation (ATT) should be added to the first temporal mode. Note that additional one-third polarization-dependent beam splitters were integrated in the gate in previous experiments \cite{gate_photon_OBrien, cphase_pdbs, cnot}, while we placed the amplitude attenuation in the stage of state preparation by Alice and Bob. This means that the states of the initial time-bin qubits are given by $\sqrt{\frac{3}{4}}\left(\sqrt{\frac{1}{3}}\left|t_1 \right\rangle+e^{i \phi} \left|t_2 \right\rangle\right)$. To implement this, we fabricated PLC interferometers equipped with additional MZ interferometers in the short arms. With these MZ interferometers of Alice and Bob as shown in the insert of Fig.~\ref{expsetup}(b), we can apply variable attenuation to the first temporal mode by adjusting the TC individually. Then the time-bin qubits are launched into the 2x2 optical switch, which is based on a lithium niobate waveguide (EO Space) \cite{takesue_switch}. By adjusting the DC bias and RF modulation signal to the PM in the switch as shown in Fig.~\ref{expsetup}(c), the 2x2 switch works as a one-third beam splitter for the $t_2$ mode and as a transparent transmission path for the $t_1$ mode. Because of the post-selection and amplitude compensation, the success probability of the C-Phase gate is 1/9 even when there are no component losses.

The output photons from ports C and D of the switch are sent to the 1-bit delay interferometers owned by Charlie and David, respectively. The photons output from the interferometers are detected by two superconducting single-photon detectors (SSPDs). The signals are used as a time interval analyzer (TIA) and the coincidences are counted by a conventional computer. The detection efficiencies of SSPD for Charlie and David are 57$\%$ and 62$\%$, respectively, and the dark count for both detectors is less than 40 cps. In order to erase the polarization distinguishability of the photon pairs, the polarization controllers (PC) are located in front of each 1-bit delay interferometer and at the input ports of the 2x2 switch, and the polarizers are placed after the outputs of the 2x2 switch. The insertion losses of the interferometers and switch are about 2.0 and 7.7 dB, respectively.

\begin{figure}[t]
\includegraphics[width=8.5cm]{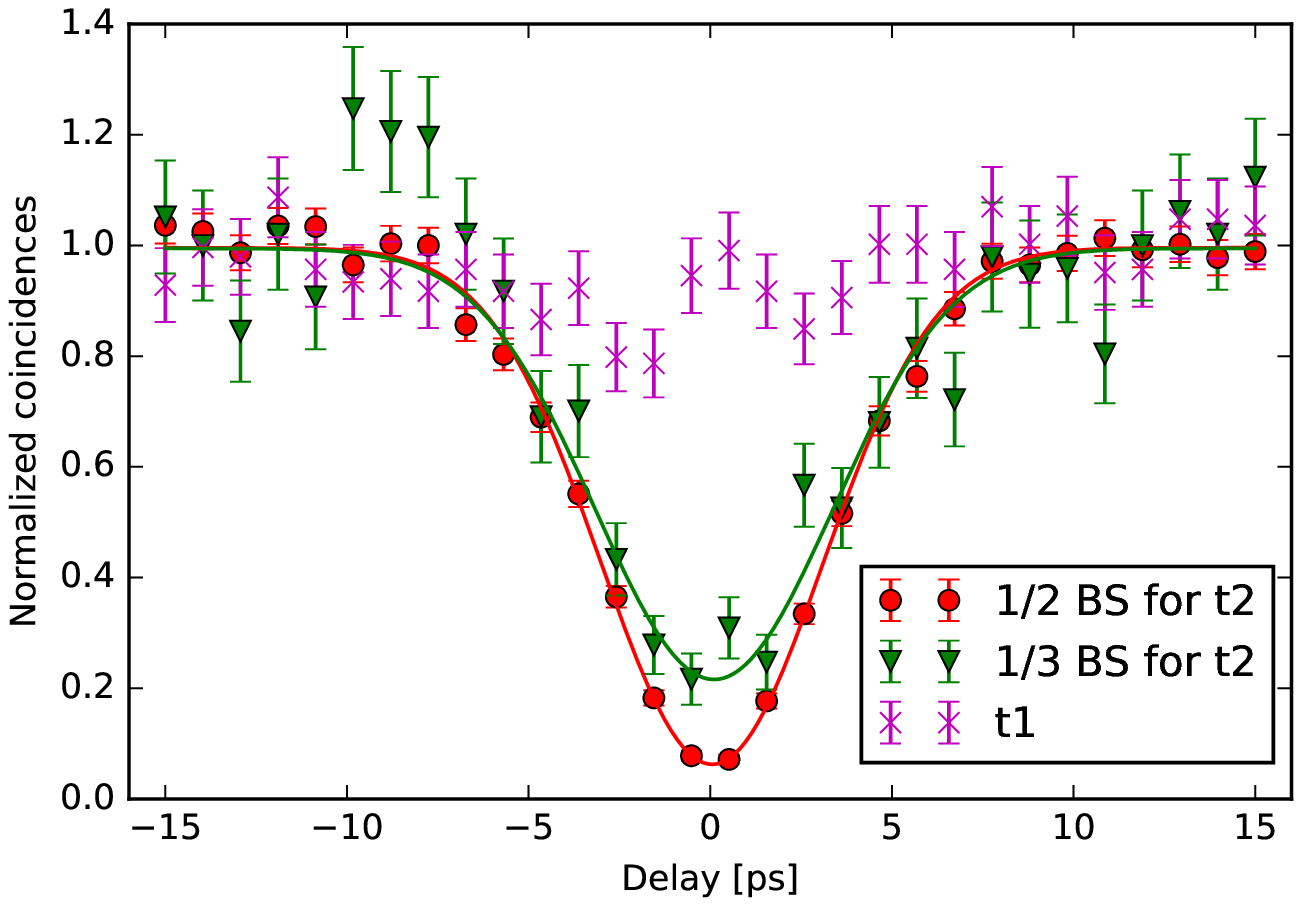}
\caption{Experimental results of HOM interference measurement. By adjusting the DC bias and modulating the RF signal to the PM, the 2x2 switch can work as a one-half (circles) and one-third (triangle) beam splitter for the second temporal mode. The visibilities are 0.937$\pm$0.002 and 0.783$\pm$0.022, which are close to the theoretically obtained visibilities of 1.0 and 0.8, respectively. The result for the first temporal mode (cross) did not show a dip, implying that the switch did not work as a beam splitter for the $t_1$ mode.}
\label{hom}
\end{figure}

\begin{figure}[htbp]  
\subfigure[Real Part] 
{  
    \begin{minipage}{3.8cm}  
    \includegraphics[scale=0.3]{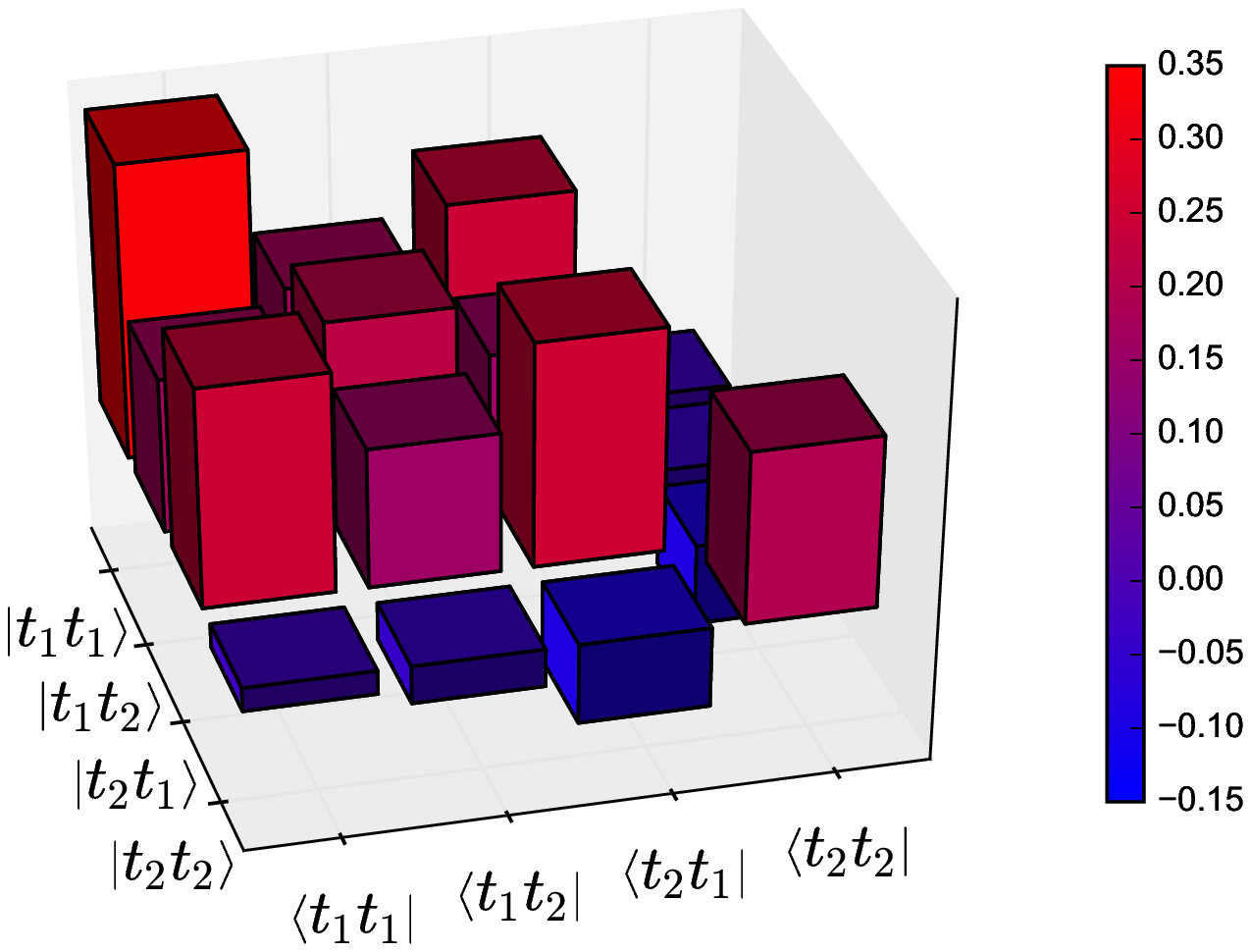}   
    \end{minipage}  
}  
\subfigure[Imaginary Part] 
{  
    \begin{minipage}{3.8cm}  
    \includegraphics[scale=0.3]{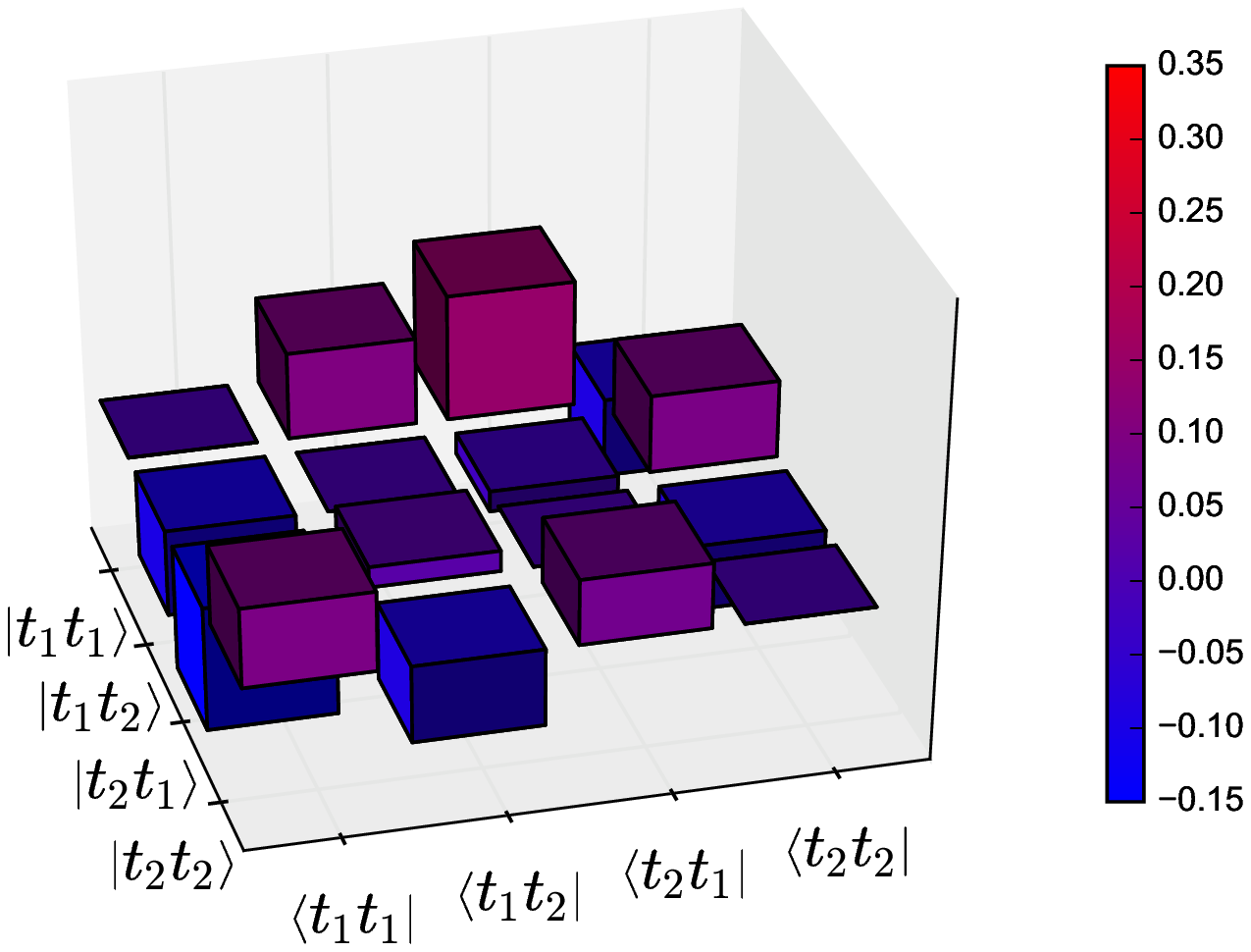}   
    \end{minipage}  
}  
\caption{(a) and (b) Real and imaginary parts of experimental density matrix determined by performing QST measurement after the C-Phase gate operation. In this measurement, Alice and Bob input the $\left|++ \right\rangle$ state by setting the phase $\phi_A=\phi_B=0$. The values of the elements of the density matrix correspond to the color bar, and the plus- and minus-sign terms are illustrated in red and blue, respectively.}  
\label{density}  
\end{figure} 

First, we confirmed that the 2x2 switch worked as time-dependent beam splitter by making a Hong-Ou-Mandel (HOM) interference measurement for time-bin qubits \cite{hom_interference}. We removed the 1-bit delay interferometers at Charlie's and David's ports and measured the coincidence counts. We adjusted the arrival times of the two time-bin qubits at the switch by changing the optical delay line (DL) placed at Bob. Here, we operated the 2x2 switch as a one-half and one-third beam splitter at the second temporal mode, while it simply passed the first temporal mode $\left|t_1 \right\rangle_C \left|t_1 \right\rangle_D$. The results are shown in Fig.~\ref{hom}. In the coincidence counts for the second temporal mode $\left|t_2 \right\rangle_C \left|t_2 \right\rangle_D$, we observed clear dips in the visibilities of 0.937$\pm$0.002 and 0.783$\pm$0.022 for the one-half and one-third beam splitter operations, respectively. These results agreed well with the corresponding theoretical visibilities, 1.0 and 0.8 for one-half and one-third beam splitters. In addition, we did not observe any dip in the result for the first temporal mode. Overall, these results confirmed that the switch could work as a time-dependent variable beam splitter as we expected. The coincidence count rate was 3 Hz outside the HOM dip, and every measurement took 40 seconds. We did not subtract accidental coincidences in any of the experiments shown in this paper.

Next, we inserted Charlie's and David's interferometers after the 2x2 switch and performed a C-Phase gate experiment. If we could implement an ideal C-Phase gate operator ($\widehat{CP}$) on Alice's and Bob's prepared photons $\left|+ \right\rangle=\frac{1}{\sqrt{2}}(\left|t_1 \right\rangle + \left|t_2 \right\rangle)$ with $\phi_A=\phi_B=0$ and $c_{1A}=c_{1B}=c_{2A}=c_{2B}=\frac{1}{\sqrt{2}}$, Charlie and David would observe a maximally time-bin entangled state, given by
\begin{eqnarray}
\begin{split}
\widehat{CP}\left|++ \right\rangle_{AB} =&  \frac{1}{2}(\left|t_1t_1 \right\rangle + \left|t_1t_2 \right\rangle + \left|t_2t_1 \right\rangle - \left|t_2t_2 \right\rangle)_{CD} \\
=&\frac{1}{\sqrt{2}}(\left|t_1+ \right\rangle+\left|t_2- \right\rangle)_{CD}.
\end{split}
\end{eqnarray} 
Thus, the density matrix of the obtained state after the C-Phase gate operation is given by
\begin{eqnarray}
\frac{1}{4}
\begin{bmatrix}
    1 & 1 & 1 & 1 \\
    1 & 1 & 1 & 1 \\
    1 & 1 & 1 & 1 \\
    1 & 1 & 1 & 1
\end{bmatrix}_{AB}
\stackrel{\widehat{CP}}{\longrightarrow} \frac{1}{4}
\begin{bmatrix}
    1 & 1 & 1 & -1 \\
    1 & 1 & 1 & -1 \\
    1 & 1 & 1 & -1 \\
    -1 & -1 & -1 & 1
\end{bmatrix}_{CD}.
\label{densitymatrix}
\end{eqnarray}
To confirm generation of an entangled state as a result of C-Phase gate operation, we performed quantum state tomography (QST) \cite{qst_pol, qst_timebin} for time-bin qubits so that we could obtain the density matrix of the output state as shown in Fig.~\ref{density}. The minus-sign terms in the density matrix of Eq.~(\ref{densitymatrix}) are clearly visible in Fig.~\ref{density}(a). The fidelity to the target entangled state was 62$\pm$7.8${\%}$. We also calculated the von Neumann entropy of 0.817, linear entropy of 0.505, and concurrence of 0.551. \cite{qst_pol, qst_timebin}

However, according to Peres \cite{peres_partialtransport}, two states are inseparable if the eigenvalues of the partial transposition of the density matrix have a negative value. The eigenvalues of the partial transposition of our density matrix were -0.204, 0.18, 0.403, and 0.621 after the C-Phase gate operation. The negative eigenvalue confirms that the time-bin qubits prepared by Alice and Bob were entangled as a result of the C-Phase gate operation implemented by the switch.

There are several points that may have decreased the fidelity of the entangled state generated by the C-Phase gate. The fluctuation of the splitting ratio of the 2x2 switch, which comes from the DC bias drift of the lithium niobate waveguide modulator, would have been the main source of the errors in the generated state \cite{cnot_theory}. Moreover, because of the large loss induced by the interferometers and the optical switch, the present experiment required a long measurement time, which increased the fluctuation of the setup further. We believe that we can obtain better fidelity by overcoming these issues \cite{cphase_shift}. In addition, the use of integrated photonics technologies will enable significant compactification of the gate function, which will lead to better stability. For example, we can integrate the function of amplitude compensation as additional intensity modulators placed in front of the 2x2 switch fabricated in a lithium niobate waveguide. 

Although we demonstrated that the C-Phase gate successfully worked for a specific input state, the present experiment does not constitute a full characterization of the quantum gate. Quantum process tomography (QPT) \cite{qpt_cnot_obrien} is now widely used for this purpose. Using QPT to analyze the gate operations requires 16 different input states, which increases the measurement time significantly. Because of the low coincidence rate caused by the relatively large component losses and the limited stability of the 2x2 switch described above, it is difficult to perform QPT with our C-Phase gate with the current setup. Therefore, it is important to reduce the component losses and improve the stability of the setup so that we can undertake QPT for more a comprehensive characterization of the gate operation.

As with the previous C-Phase gates based on post-selection \cite{cphase_theory, cphase_pdbs}, the limited success probability will constrict the application of these gates to systems with a relatively small number of qubits. For example, such probabilistic quantum gates could be useful for demonstrating a quantum communication system based on quantum error correction \cite{Bill_repeater}.

We would like to note that a C-NOT gate can be performed by applying a Hadamard transform on the target time-bin qubit before and after the C-Phase gate \cite{cnot}. In addition, we can tune the amount of phase shift given to the time-bin qubits by changing the amplitude of the modulation signal to the switch \cite{cphase_shift}. 

When we use the proposed gate in a quantum network over optical fiber, precise adjustment of the path lengths in front of the 2x2 switch is not a trivial issue. However, such path-length matching in a fiber network has been successfully demonstrated in several long-distance quantum teleportation experiments \cite{teleportation_network_JWPan_NP_2016, teleportation_network_WTittel_NP_2016}, in which active feedback control based on HOM interference measurement was implemented, together with the sharing of the time reference between nodes enabled by the use of classical channels. These techniques can be applied to deploy our quantum gate in a real fiber network.

In summary, we demonstrated a C-Phase gate for time-bin qubits by using a 2x2 switch as a beam splitter. By adjusting the DC and RF signal, the optical switch can work as a time-dependent beam splitter with different splitting ratios for different temporal modes. Here, the 2x2 switch was operated as a one-third beam splitter of the $t_2$ mode that passes the $t_1$ mode. The HOM experiment showed that the visibility was 0.78$\pm$0.02 for the $t_2$ mode and there was no dip for the $t_1$ mode. By performing QST, an examination of the density matrix showed that the C-Phase gate successfully entangled the time-bin states prepared by Alice and Bob, and the fidelity was 62$\pm$7.8${\%}$. 

We thank William J. Munro for fruitful discussions.


\begin{thebibliography}{999}
\bibitem{QST_superconducting_IBM} M. Steffen, M. Ansmann, R. C. Bialczak, N. Katz, E. Lucero, R. McDermott, M. Neeley, E. M. Weig, A. N. Cleland, and J. M. Martinis, Science 313, 1423 (2006).

\bibitem{cnot_gate_superconducting} J. H. Plantenberg, P. C. de Groot, C. J. P. M. Harmans and J. E. Mooij, Nature 447, 836-839 (2007).

\bibitem{gate_superconducting} L. DiCarlo, J. M. Chow, J. M. Gambetta, Lev S. Bishop, B. R. Johnson, D. I. Schuster, J. Majer, A. Blais, L. Frunzio, S. M. Girvin, and R. J. Schoelkopf, Nature 460, 240-244 (2009).

\bibitem{steane_iontrap} A. Steane, Appl. Phys. B 64, 623-642 (1997). 
\bibitem{gate_ion_trap} M. Riebe, K. Kim, P. Schindler, T. Monz, P. O. Schmidt, T. K. K\"{o}rber, W. H\"{a}nsel, Phys. Rev. Lett. 97, 220407 (2006).

\bibitem{Hanson_dot_RMP_2007} R. Hanson, L. P. Kouwenhoven, J. R. Petta, S. Tarucha, and L. M. K. Vandersypen, Rev. Mod. Phys. 79, 1217 (2007).

\bibitem{Zwanenburg_dot_RMP_2007} F. A. Zwanenburg, A. S. Dzurak, A. Morello, M. Y. Simmons, L. C. L. Hollenberg, G. Klimeck, S. Rogge, S. N. Coppersmith, and M. A. Eriksson, Rev. Mod. Phys. 85, 961 (2013).

\bibitem{Li_Qdot_gate} X. Li, Y. Wu, D. Steel, D. Gammon, T. H. Stievater, D. S. Katzer, D. Park, C. Piermarocchi, and L. J. Sham, Science 313, 1423 (2006).

\bibitem{Nielsen_Chuang} M. A. Nielsen and I. L. Chuang, Quantum Computation and Quantum Information (Cambridge: Cambridge University Press, 2000), Chap. 4.



\bibitem{KLM} E. Knill, R. Laflamme and G. J. Milburn, Nature 409, 46-52 (2001).



\bibitem{cnot_theory} T. C. Ralph, N. K. Langford, T. B. Bell, and A. G. White, Phys. Rev. A 65, 062324 (2002).


\bibitem{cphase_theory} H. F. Hofmann, and S. Takeuchi, Phys. Rev. A 66, 024308 (2002).


\bibitem{GHZ_LOQC_2015} M. Gimeno-Segovia, P. Shadbolt, D. E. Browne, and T. Rudolph, Phys. Rev. Lett. 115, 020502 (2015).

\bibitem{gate_photon_OBrien} J. L. O'Brien, G. J. Pryde, A. G. White, T. C. Ralph, and D. Branning, Nature 426, 264-267 (2003).

\bibitem{cphase_pdbs} N. Kiesel, C. Schmid, U. Weber, R. Ursin, and H. Weinfurter, Phys. Rev. Lett. 95, 210505 (2005).

\bibitem{cnot_Pittman_2003} T. B. Pittman,M. J. Fitch, B. C Jacobs, and J. D. Franson, Phys. Rev. A 68, 032316 (2003).

\bibitem{gate_Bell} N. K. Langford, T. J. Weinhold, R. Prevedel, K. J. Resch, A. Gilchrist, J. L. O'Brien, G. J. Pryde, and A. G. White, Phys. Rev. Lett. 95, 210504 (2005).

\bibitem{cnot} R. Okamoto, H. F. Hofmann, S. Takeuchi, and K. Sasaki, Phys. Rev. Lett. 95, 210506 (2005).


\bibitem{kok_RMP} P. Kok, W. J. Munro, K. Nemoto, T. C. Ralph, J. P. Dowling, and G. J. Milburn, Rev. Mod. Phys. 79, 135 (2007).

\bibitem{Bill_repeater} W. J. Munro, A. M. Stephens, S. J. Devitt, K. A. Harrison, and K. Nemoto, Nature Photon. 6, 777-781 (2012).

\bibitem{time-energy_Brendel_1999} J. Brendel, N. Gisin, W. Tittel, and H. Zbinden, Phys. Rev. Lett. 82, 2594 (1999).

\bibitem{time-energy_Gisin_2007} N. Gisin and R. Thew, Nat. Photonics 1, 165 (2007).

\bibitem{Honjo_timebin_QKD} T. Honjo, S. W. Nam, H. Takesue, Q. Zhang, H. Kamada, Y. Nishida, O. Tadanaga, M. Asobe, B. Baek, R. Hadfield, S. Miki, M. Fujiwara, M. Sasaki, Z. Wang, K. Inoue, and Y. Yamamoto, Opt. Express 16, 19118 (2008).


\bibitem{takesue_switch} H. Takesue, Phys. Rev. A 89, 062328 (2014).

\bibitem{takesue_plc} H. Takesue, and K. Inoue, Phys. Rev. A 72, 041804(R) (2005).
\bibitem{hom_interference} C. K. Hong, Z. Y. Ou, and L. Mandel, Phys. Rev. Lett. 59, 2044 (1987).

\bibitem{qst_pol} D. F. V. James, P. G. Kwiat, W. J. Munro, and A. G. White, Phys. Rev. A 64, 052312 (2001).
\bibitem{qst_timebin} H. Takesue and Y. Noguchi, Opt. Express 17, 10976-10989 (2009).

\bibitem{peres_partialtransport} A. Peres, Phys. Rev. Lett. 77, 1413 (1996).

\bibitem{cphase_shift} K. Lemr, A. \u{C}ernoch, J. Soubusta, K. Kieling, J. Eisert, and M. Du\u{s}ek, Phys. Rev. Lett. 106, 013602 (2011).

\bibitem{qpt_cnot_obrien} J. L. O'Brien, G. J. Pryde, A. Gilchrist, D. F. V. James, N. K. Langford, T. C. Ralph, and A. G. White, Phys. Rev. Lett. 93, 080502 (2004).

\bibitem{teleportation_network_JWPan_NP_2016} Q.-C. Sun, Y.-L. Mao, S.-J. Chen, W. Zhang, Y.-F. Jiang, Y.-B. Zhang, Q.-J. Zhang, S. Miki, T. Yamashita, H. Terai, X. Jiang, T.-Y. Chen, L.-X. You, X.-F. Chen, Z. Wang, J.-Y. Fan, Q. Zhang, and J.-W. Pan, Nat. Photonics 10, 671 (2016).

\bibitem{teleportation_network_WTittel_NP_2016} R. Valivarth, M. G. Puigibert, Q. Zhou, G. H. Aguilar, V. B. Verma, F. Marsili, M. D. Shaw, W. Nam, D. Oblak, and W. Tittel, Nat. Photonics 10, 676 (2016).

\end{thebibliography}
\end{document}